\documentclass[12pt]{article}
\usepackage{epsfig}
\def\be{\begin{equation}}
\def\ee{\end{equation}}
\def\bea{\begin{eqnarray}}
\def\eea{\end{eqnarray}}
\usepackage{graphicx}

\catcode`\@=11
\def\lsim{\mathrel{\mathpalette\@versim<}}
\def\gsim{\mathrel{\mathpalette\@versim>}}
\def\@versim#1#2{\vcenter{\offinterlineskip
\ialign{$\m@th#1\hfil##\hfil$\crcr#2\crcr\sim\crcr } }}
\catcode`\@=12
\usepackage{axodraw}

\catcode`@=12
\begin{document}
\thispagestyle{empty}
\begin{flushright}
UCRHEP-T527\\
April 2013\
\end{flushright}
\vspace{0.6in}
\begin{center}
{\LARGE \bf Neutrino Mixing and Geometric $CP$ Violation\\
with $\Delta(27)$ Symmetry\\}
\vspace{1.2in}
{\bf Ernest Ma\\}
\vspace{0.2in}
{\sl Department of Physics and Astronomy, University of California,\\
Riverside, California 92521, USA\\}
\end{center}
\vspace{1.2in}
\begin{abstract}\
Predictive spontaneous $CP$ violation is possible if it is obtained 
geometrically through a non-Abelian discrete symmetry. 
I propose such a model of neutrino mass and mixing based on 
$\Delta(27)$.  
\end{abstract}

\newpage
\baselineskip 24pt

Since the experimental determination of nonzero $\theta_{13}$ in neutrino 
oscillations, the next big question in neutrino physics is $CP$ violation. 
Theoretically, this should be understood together with the mixing angles 
themselves.  Whereas non-Abelian discrete symmetries (the 
first~\cite{mr01,m02,bmv03,m04} of which was $A_4$) are useful in 
obtaining tribimaximal mixing~\cite{hps02} which requires $\theta_{13}=0$ 
and no $CP$ violation, the data now require either a modification or 
a new approach.  In the former, $CP$ violation may be incorporated by 
allowing nonzero $\theta_{13}$ and complex Yukawa couplings.  A simple 
example is a variation~\cite{im12} of the original $A_4$ model~\cite{m04} 
for tribimaximal mixing.  In the latter, the discrete symmetry may be extended 
to include generalized $CP$ transformations~\cite{gl04}, which in the 
case~\cite{mn12} of $S_4$ could lead to maximal $CP$ violation as well as 
maximal $\theta_{23}$.  Another possible approach in this category is 
spontaneous geometric $CP$ violation~\cite{bgg84} using $\Delta(27)$, which 
has recently been applied~\cite{bdl12} successfully to the quark sector.  
This paper deals with the lepton sector~\cite{m06-1,dkr07,m08} and how it 
may be related~\cite{m06} to dark matter.

The non-Abelian discrete symmetry $\Delta(27)$ has 27 elements, with nine 
one-dimensional irreducible representations $\underline{1}_i$ ($i = 1,...,9$) 
and two three-dimensional ones $\underline{3}$ and $\underline{3}^*$. 
Its $11 \times 11$ character table as well as the 27 defining $3 \times 3$ 
matrices of its $\underline{3}$ representation are given in Ref.~\cite{m06-1}.
The group multiplcation rules are
\begin{equation}
\underline{3} \times \underline{3} = \underline{3^*} + \underline{3}^* 
+ \underline{3}^*, ~~~~~ \underline{3} \times \underline{3}^* = \sum_{i=1}^9 
\underline{1}_i.
\end{equation}
The important property to notice is that  $\underline{3} \times \underline{3}
\times \underline{3}$ has three invariants: 
$123 + 231 + 312 - 213 - 321 - 132$ [which is also invariant under $SU(3)$], 
$123 + 231 + 312 + 213 + 321 + 132$ [which is also invariant under $A_4$], 
and $111 + 222 +333$.

In this paper, the assignments of the lepton and Higgs fields are different 
from previous studies~\cite{m06-1,dkr07,m08}, with the new requirement 
that $CP$ be spontaneously broken geometrically~\cite{bgg84,bdl12}.
Let
\begin{equation}
\pmatrix{\nu \cr l}_i \sim \underline{3}, ~~~ l^c_i \sim \underline{1}_1, 
\underline{1}_2, \underline{1}_2, ~~~ \pmatrix{\phi^+ \cr \phi^0}_i \sim 
\underline{3}.
\end{equation}
Using the decomposition of $\underline{3} \times \underline{3}^*$ and 
$\langle \phi^0_i \rangle = v_i$, the charged-lepton mass matrix is 
given by
\begin{eqnarray}
{\cal M}_l &=& \pmatrix{f_e v_1^* & f_\mu v_1^* & f_\tau v_1^* \cr 
f_e v_2^* & f_\mu \omega^2 v_2^* & f_\tau \omega v_2^* \cr 
f_e v_3^* & f_\mu \omega v_3^* & f_\tau \omega^2 v_3^*} 
= \pmatrix{v_1^* & 0 & 0 \cr 0 & v_2^* & 0 \cr 0 & 0 & v_3^*} 
\pmatrix{1 & 1 & 1 \cr 1 & \omega^2 & \omega \cr 1 & \omega & \omega^2} 
\pmatrix{f_e & 0 & 0 \cr 0 & f_\mu & 0 \cr 0 & 0 & f_\tau},  
\end{eqnarray}
where $\omega = \exp (2 \pi i/3) = -1/2 + i\sqrt{3}/2$.  This ${\cal M}_l$ 
is identical in form to that of the original $A_4$ model of Ref.~\cite{mr01}. 
The new feature here is that $CP$ conservation is imposed on the Lagrangian 
(so that all the Yukawa couplings are real) but it is spontaneously broken 
by the vacuum, i.e.~\cite{bgg84,bdl12}
\begin{equation}
(v_1, v_2, v_3) = v (\omega, 1, 1).
\end{equation}
Hence
\begin{equation}
{\cal M}_l = \pmatrix{\omega^2 & 0 & 0 \cr 0 & 1 & 0 \cr 0 & 0 & 1} 
{1 \over \sqrt{3}} \pmatrix{1 & 1 & 1 \cr 1 & \omega^2 &  \omega \cr 1 & 
\omega & \omega^2} \pmatrix{m_e & 0 & 0 \cr 0 & m_\mu & 0 \cr 0 & 0 & m_\tau},
\end{equation}
where $m_e = \sqrt{3} f_e v$, etc.

For the neutrino mass matrix, three Higgs doublets
\begin{equation}
\pmatrix{\zeta^+ \cr \zeta^0}_i \sim \underline{1}_1, \underline{1}_2, 
\underline{1}_3
\end{equation}
are added so that the dimension-five operator $\Lambda^{-1} (\nu \nu \phi^0) 
\zeta^0$ for the $3 \times 3$ Majorana neutrino mass matrix has six 
invariants, i.e. 
\begin{equation}
{\cal M}_\nu = \pmatrix{\omega(f_1+f_2+f_3) & f_4 + \omega f_5 + \omega^2 f_6 
& f_4 + \omega^2 f_5 + \omega f_6 \cr f_4 + \omega f_5 + \omega^2 f_6 & 
f_1 + \omega^2 f_2 + \omega f_3 & \omega(f_4 + f_5 + f_6) \cr f_4 + \omega^2 
f_5 + \omega f_6 & \omega(f_4 + f_5 + f_6) & f_1 + \omega f_2 + \omega^2 f_3},
\end{equation}
where $\Lambda^{-1} v \langle \zeta^0_i \rangle$ have been absorbed into the 
definitions of the $f$ parameters.

Using Eq.~(5), the neutrino mass matrix in the tribimaximal basis is now 
given by
\begin{eqnarray}
{\cal M}_\nu^{(1,2,3)} &=& \pmatrix{0 & 1/\sqrt{2} & 1/\sqrt{2} \cr \omega & 
0 & 0 \cr 0 & -i/\sqrt{2} & i/\sqrt{2}} {\cal M}_\nu \pmatrix{0 & \omega & 
0 \cr 1/\sqrt{2} & 0 & -i/\sqrt{2} \cr 1/\sqrt{2} & 0 & i/\sqrt{2}} \nonumber 
\\ &=& \pmatrix{\omega d + b & \omega e & c \cr \omega e & a & \omega f \cr 
c & \omega f & \omega d -b},
\end{eqnarray}
where $a = f_1 + f_2 + f_3$, $b = f_1 - (f_2+f_3)/2$, 
$c = \sqrt{3}(f_3-f_2)/2$, $d = f_4 + f_5 + f_6$, 
$e = \sqrt{2} f_4 - (f_5+f_6)/\sqrt{2}$, $f = \sqrt{3}(f_5-f_6)/\sqrt{2}$.
The tribimaximal limit, i.e.
\begin{equation}
U_{l\nu} = \pmatrix{\sqrt{2/3} & 1/\sqrt{3} & 0 \cr -1/\sqrt{6} & 1/\sqrt{3} 
& -1/\sqrt{2} \cr -1/\sqrt{6} & 1/\sqrt{3} & 1/\sqrt{2}}
\end{equation}
is reached for $c=e=f=0$.  To lowest order, $c \neq 0$ implies 
$\tan^2 \theta_{12} > 0.5$ and $\theta_{13} \neq 0$; $e \neq 0$ implies 
$\tan^2 \theta_{12}$ can be greater or less than 1/2 and $\theta_{13}=0$; 
$f \neq 0$ implies $\tan^2 \theta_{12} < 1/2$ and $\theta_{13} \neq 0$.
Given that data prefer the last choice, it will be assumed from now on 
that $c$ and $e$ are negligible and only nonzero $f$ is considered.  
The immediate consequence~\cite{im12} of this is that $\theta_{12}$ 
and $\theta_{13}$ are related, and that given $\theta_{13}$ and $\theta_{23}$, 
$|\tan \delta_{CP}|$ is determined. 

Since $c=e=0$ has been assumed, ${\cal M}_\nu^{(1,2,3)}$ is diagonalized 
by
\begin{equation}
\pmatrix{m_2 & 0 \cr 0 & m_3} = \pmatrix{\cos \theta & \sin \theta e^{i \phi} 
\cr -\sin \theta e^{-i \phi} & \cos \theta} \pmatrix{a & \omega f \cr \omega 
f & \omega d - b} \pmatrix{\cos \theta & -\sin \theta e^{-i \phi} 
\cr \sin \theta e^{i \phi} & \cos \theta}. 
\end{equation}
Since $a,b,d,f$ are real, this implies
\begin{equation}
\tan \phi = {\sqrt{3}(a+b) \over a-b-2d}, ~~~~~ \tan 2 \theta = 
{4f \sqrt{a^2 + b^2 + d^2 + ab - ad + bd} \over b^2 - 2a^2 + 2d^2 - ab}.
\end{equation}
With this structure, $|\sin \theta_{13}| = |\sin \theta|/\sqrt{3}$, which 
implies
\begin{equation}
\tan^2 \theta_{12} = {1 - 3 \sin^2 \theta_{13} \over 2},
\end{equation}
which agrees very well~\cite{im12} with data.  As for the phase $\phi$, 
it is given by the condition
\begin{equation}
\tan^2 \theta_{23} = {\left( 1 - {\sqrt{2} \sin \theta_{13} \cos \phi \over 
\sqrt{1-3\sin^2 \theta_{13}}} \right)^2 + {2 \sin^2 \theta_{13} \sin^2 \phi 
\over 1-3 \sin^2 \theta_{13}} \over 
\left( 1 + {\sqrt{2} \sin \theta_{13} \cos \phi \over 
\sqrt{1-3\sin^2 \theta_{13}}} \right)^2 + {2 \sin^2 \theta_{13} \sin^2 \phi 
\over 1-3 \sin^2 \theta_{13}}}.
\end{equation}

Since $m_2^2$ and $m_3^2$ are corrected by terms proportional to $f^2$ 
which are small, the following approximation for the neutrino masses 
is valid for the analysis below, i.e.
\begin{equation}
m_1 = \sqrt{b^2 -db + d^2},  ~~~ m_2 = |a|, ~~~ m_3 = \sqrt{b^2 +db + d^2}.
\end{equation}
Hence $2bd = \pm |\Delta m^2_{32}| \equiv \pm \Delta$ for normal (inverted) 
ordering of neutrino masses.  Since $\Delta m^2_{21} << \Delta$, $m_1 
\simeq m_2$ will be also assumed below.

Let $\Delta = 2.35 \times 10^{-3}$ eV$^2$, which is the central value from the 
2012 PDG compilation, then using $d = \pm \Delta/2b$ and 
$a = \pm \sqrt{b^2 - bd + d^2}$, this model has the prediction
\begin{eqnarray}
&& \sum m > (2 + \sqrt{3}) \sqrt{\Delta \over 2} = 0.13~{\rm eV}
~{\rm for~normal ~ordering}, \\
&& \sum m > (2\sqrt{3} + 1) \sqrt{\Delta \over 2} = 0.15~{\rm eV}
~{\rm for~inverted ~ordering}.
\end{eqnarray}
Using the latest Planck result~\cite{planck13} that $\sum m < 0.23$ eV, 
the range of values for $b$ is also obtained:
\begin{eqnarray}
&& 0.015 < b < 0.078~{\rm eV}~{\rm for~normal~ordering}, \\
&& 0.016 < b < 0.073~{\rm eV}~{\rm for~inverted~ordering}.
\end{eqnarray}
Using Eq.~(13) for $\sin^2 2 \theta_{23} > 0.92$ and $\sin^2 2 \theta_{13} 
\simeq 0.1$, the constraint
\begin{equation}
|\tan \phi| > 1, ~~{\rm or}~~ |\sin \phi| > 1/\sqrt{2}
\end{equation}
is obtained.  Using Eq.~(11), this restricts $a > 0$ for normal ordering, 
and $a > 0$ with $b > 0.02$ or $a < 0$ with $b < 0.04$ for inverted 
ordering of neutrino masses.

The invariant $CP$ violating 
parameter $J_{CP} = Im(U_{\mu 3} U^*_{e3} U_{e2} U^*_{\mu 2})$ is simply 
given in this model by
\begin{equation}
J_{CP} = {\sin \theta_{13} \sqrt{1 - 3 \sin^2 \theta_{13}} \sin \phi \over 
3 \sqrt{2}}.
\end{equation}
Using $\sin \theta_{13} \simeq 0.16$ and $|\sin \phi| > 1/\sqrt{2}$, 
the allowed range
\begin{equation}
0.026 < |J_{CP}| < 0.036
\end{equation}
is thus obtained.  As for the effective neutrino mass in neutrinoless 
double beta decay, its allowed range is approximately given by 
\begin{equation}
0.03 < m_{ee} < 0.07~{\rm eV}.
\end{equation}
Thus this model has two very specific predictions: (1) $|J_{CP}|$ is 
between 0.026 and 0.036, and (2) $m_{ee}$ is between 0.03 and 0.07 eV.

The dimension-five operator~\cite{w79} for Majorana neutrino mass considered 
in the above may be implemented~\cite{m06} in one loop, with 
dark matter ($Z_2$ odd) in the loop.
\begin{figure}[htb]
\begin{center}
\begin{picture}(260,120)(0,0)
\ArrowLine(60,10)(90,10)
\ArrowLine(130,10)(90,10)
\ArrowLine(130,10)(170,10)
\ArrowLine(200,10)(170,10)
\DashArrowArc(130,10)(40,90,180)5
\DashArrowArcn(130,10)(40,90,0)5
\DashArrowLine(100,80)(130,50)5
\DashArrowLine(160,80)(130,50)5
\Text(75,0)[]{\large $\nu$}
\Text(185,0)[]{\large $\nu$}
\Text(110,0)[]{\large $N$}
\Text(150,0)[]{\large $N$}
\Text(100,52)[]{\large $\eta^0$}
\Text(160,52)[]{\large $\eta^0$}
\Text(95,90)[]{\large $\langle \phi^0 \rangle$}
\Text(165,90)[]{\large $\langle \phi^0 \rangle$}
\Text(130,10)[]{\Large $\times$}
\end{picture}
\end{center}
\caption{One-loop generation of scotogenic Majorana neutrino mass.}
\end{figure}
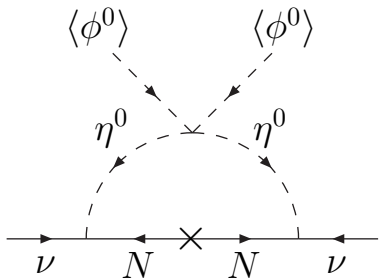
This mechanism has been called ``scotogenic'', from the Greek ``scotos'' 
meaning darkness.  Because of the allowed $(\lambda_5/2) (\Phi^\dagger \eta)^2 
+ H.c.$ interaction, $\eta^0 = (\eta_R + i \eta_I)/\sqrt{2}$ is split so 
that $m_R \neq m_I$.  The diagram of Fig.~1 can be computed 
exactly~\cite{m06}, i.e.
\begin{equation}
({\cal M}_\nu)_{ij} = \sum_k {h_{ik} h_{jk} M_k \over 16 \pi^2} 
\left[ {m_R^2 \over m_R^2 - M_k^2} \ln {m_R^2 \over M_k^2} - 
{m_I^2 \over m_I^2 - M_k^2} \ln {m_I^2 \over M_k^2} \right].
\end{equation} 
A good dark-matter candidate is $\eta_R$ as first pointed out in 
Ref.~\cite{m06}, whereas its stabilty was already anticipated in 
Ref.~\cite{dm78}.  
It was subsequently proposed by itself as dark matter in Ref.~\cite{bhr06} 
(to render the standard-model Higgs boson very heavy, which is now ruled 
out by data) and studied in detail in Ref.~\cite{lnot07}.  The $\eta$ 
doublet has become known as the ``inert'' Higgs doublet, but it does have 
gauge and scalar interactions even if it is the sole addition to the 
standard model.   In principle, the lightest $N$ is also a possible 
dark-matter candidate~\cite{kms06}, but its mass and couplings may be  
severely restricted by the experimental limit on $\mu \to e \gamma$ decay, 
unless a symmetry exists to suppress it, which is possible in this case.

To accommodate the $\Delta(27)$ symmetry, the external $\phi^0 \phi^0$ 
lines are replaced by $\phi^0_i \zeta^0_j$, and the internal $\eta^0$ 
($N$) lines are replaced by $\eta^0_i,~N_i \sim \underline{3}$ on one side, 
and $\eta^0 \sim \underline{1}$, $N_i \sim \underline{3}^*$ on the other.

In conclusion, a special mechanism of $CP$ violation has been implemented in 
a complete model of charged-lepton and neutrino masses and mixing, using 
the non-Abelian discrete symmetry $\Delta(27)$.  The Lagrangian is 
required to conserve $CP$ resulting in real Yukawa couplings, but the 
Higgs vacuum breaks $CP$ spontaneously and geometrically.  The resulting 
model has some very specific predictions, as given by Eqs.~(12) to (22).

\noindent \underline{Acknowledgment}~:~ I thank the hospitality of 
the Institute of Advanced Studies, Nanyang Technological University, 
Singapore, and the Institute of Physics, Academia Sinica, Taipei, Taiwan. 
This work is supported in part by the U.~S.~Department of Energy under 
Grant No.~DE-AC02-06CH11357.

\bibliographystyle{unsrt}

\begin{thebibliography}{99}
\bibitem{mr01} E. Ma and G. Rajasekaran, Phys. Rev. {\bf D64}, 113012 (2001).
\bibitem{m02} E. Ma, Mod. Phys. Lett. {\bf A17}, 2361 (2002).
\bibitem{bmv03} K. S. Babu, E. Ma, and J. W. F. Valle, Phys. Lett. {\bf B552}, 
207 (2003).
\bibitem{m04} E. Ma, Phys. Rev. {\bf D70}, 031901 (2004).
\bibitem{hps02} P. F. Harrison, D. H. Perkins, and W. G. Scott, Phys. Lett. 
{\bf B530}, 167 (2002).
\bibitem{im12} H. Ishimori and E. Ma, Phys. Rev. {\bf D86}, 045030 (2012). 
\bibitem{gl04} W. Grimus and L. Lavoura, Phys. Lett. {\bf B579}, 113 (2004).
\bibitem{mn12} R. N. Mohapatra and C. C. Nishi, Phys. Rev. {\bf D86}, 073007 
(2012)
\bibitem{bgg84} G. C. Branco, J. M. Gerard, and W. Grimus, Phys. Lett. 
{\bf B136}, 383 (1984).
\bibitem{bdl12} G. Bhattacharyya, I. de Medeiros Varzielas, and P. Leser, 
Phys. Rev. Lett. {\bf 109}, 241603 (2012).
\bibitem{m06-1} E. Ma, Mod. Phys. Lett. {\bf A21}, 1917 (2006).
\bibitem{dkr07} I. de Medeiros Varzielas, S. F. King, and G. G. Ross, 
Phys. Lett. {\bf B648}, 201 (2007).
\bibitem{m08} E. Ma, Phys. Lett. {\bf B660}, 505 (2008).
\bibitem{m06} E. Ma, Phys. Rev. {\bf D73}, 077301 (2006).
\bibitem{planck13} Planck Collaboration, P. A. R. Ade {\it et al.}, 
arXiv:1303.5076 [astro-ph/CO].
\bibitem{w79} S. Weinberg, Phys. Rev. Lett. {\bf 43}, 1566 (1979). 
\bibitem{dm78} N. G. Deshpande and E. Ma, Phys. Rev. {\bf D18}, 2574 (1978).
\bibitem{bhr06} R. Barbieri, L. J. Hall, and V. S. Rychkov, Phys. Rev. 
{\bf D74}, 015007 (2006).
\bibitem{lnot07} L. Lopez Honorez, E. Nezri, J. F. Oliver, and M. H. G. 
Tytgat, JCAP {\bf 0702}, 028 (2007).
\bibitem{kms06} J. Kubo, E. Ma, and D. Suematsu, Phys. Lett. {\bf B642}, 
18 (2006).








\end{thebibliography}

\end{document}